\documentclass[twocolumn, journal,10pt]{IEEEtran}

\usepackage{amssymb}
\usepackage{times,amsmath,epsfig,algorithmic,algorithm,color,fixltx2e}

\ifCLASSINFOpdf
\else
\fi
\begin{document}
%
\title{Adaptive Decision Feedback Detection with Parallel Interference Cancellation and Constellation Constraints for Multi-Antenna Systems}
%
%
%

\author{Peng~Li,~Rodrigo~C.~de~Lamare,~Jingjing~Liu
\thanks{Part of this paper was presented at ICASSP2012. Peng~Li, Jingjing Liu and Rodrigo~C.~de.~Lamare are with the Department
of Electronics, The University of York, England,
UK, YO10 5DD e-mail: (pl534,rcdl500,jl622)@ohm.york.ac.uk.}
}

%
%

\markboth{Draft paper for {\bf{IET Communications}}, April~2012}%
{Shell \MakeLowercase{\textit{et al.}}: Bare Demo of IEEEtran.cls for Journals}

%



\maketitle

\begin{abstract}

In this paper, a novel low-complexity adaptive decision feedback
detection with parallel {  decision feedback} and constellation
constraints (P-DFCC) is proposed for multiuser MIMO systems. We
propose a constrained constellation map which introduces a number of
selected points served as the feedback candidates for interference
cancellation. By introducing a reliability checking, a higher degree
of freedom is introduced to refine the unreliable estimates. The
P-DFCC is followed by {  an adaptive receive filter} to estimate the
transmitted symbol. In order to reduce the complexity of computing
the filters with time-varying MIMO channels, an adaptive recursive
least squares (RLS) algorithm is employed in the proposed P-DFCC
scheme. An iterative detection and decoding (Turbo) scheme is
considered with the proposed P-DFCC algorithm. Simulations show that
the proposed technique has a complexity comparable to { the
conventional parallel decision feedback detector} while it obtains a
performance close to the maximum likelihood detector at {  a low to
medium SNR range.}
\end{abstract}


%
\IEEEpeerreviewmaketitle

\section{Introduction}
%
%
%
%

\IEEEPARstart{M}{ulti}-user detection (MUD) \cite{verdu} algorithms
have shown that they can be applied to the uplink of 3G and next
generation multi-antenna communication systems. MUD can also be
applied to spatially multiplexed multi-input multi-output (MIMO)
wireless communication systems to form a spatial division multiple
access (SDMA) scheme. In such systems, multiple users are operated
within the same frequency band simultaneously and the spatial
dimension is exploited which can significantly increase the
bandwidth efficiency. In order to successfully restore the signals
from the received signal combination, pre-coding \cite{NG05},
\cite{LH05} and decoding \cite{delamare_spa}, \cite{CL09} techniques
are developed at the transmitter side and receiver side
respectively. Due to the fact that for a multiple access uplink
scenario, it is difficult for each user equipment (UE) to know the
channel state information (CSI) of others, in this paper, we focus
on the decoding part.


Several detection techniques have been developed for use at the
receiver to suppress the multi-access interference (MAI), recover
the simultaneously transmitted signals and increase the throughput
for the served UEs \cite{delamare_spa}. The optimal maximum
likelihood detection (MLD) \cite{verdu} scheme has exponential
complexity with the number of data streams and the modulation level,
which is impractical for systems even with a moderate number of UEs.
The cost effective ML solutions such as sphere decoders (SD)
\cite{viterbo} \cite{HB03} approach the optimal performance with
reduced complexity \cite{vikalo}, however, they still have a lower
bound complexity which is polynomial or exponential depending on the
number of UEs as well as the signal-to-noise ratio (SNR)
\cite{FL11}. In order to avoid the high complexity of ML or near-ML
detectors, {  linear detectors which are based on minimum mean
square error (MMSE) or zero-forcing receiver filters have been
investigated}. Generally, linear detectors experience a performance
loss and achieve a lower capacity. A decision feedback receiver {
with successive decision feedback (S-DF) \cite{LL11_TWC} or with
parallel decision feedback (P-DF)} \cite{mdfpic} can be employed to
achieve a higher capacity. {  These DF receiver structures
\cite{ginis}, \cite{lamareMBER},\cite{Foschini3},\cite{delamare_spa}
are preferred as they offer an attractive performance and complexity
trade-off, which is usually a key concern in multiple access
systems.}

The S/P-DF architectures are able to provide high spectral
efficiencies when multiple transmit antennas are deployed
\cite{LH05}. However, the application to systems with time-varying
channels is difficult due to the excessive computational load for
updating the receive filter coefficients and tracking the channel
\cite{sun}. {  The estimation of the receive filter weights and the
CSI requires matrix inversions and other operations that lead to a
significant number of computations.}


As an alternative, the training aided adaptive techniques may be
deployed for multiuser systems in time-varying channels \cite{choi}.
Adaptive algorithms can be used to track the channels and to avoid
excessive computations when the channels are varying. In
\cite{choi}, the authors developed a low-complexity data-aided
adaptive technique for detecting the time-varying channels based on
the GDF \cite{ginis} structure, the weight vectors are updated using
the recursive least squares (RLS) based algorithm. The multiple
access interference introduced by spatial multiplexing can be
suppressed in a serial or parallel manner and the transmitted
symbols are estimated at each stage. { Despite its many benefits,
there is a large performance loss when one compares the performance
of a DF based receiver with that of the optimal detector.} This is
due to the fact that (1) { the DF structure can not provide the full
receive diversity order achieved by the optimal MLD in spatially
multiplexed systems.} (2) { The average performance of S/P-DF} is
dominated by the data stream with the lowest SINR and the effect of
error propagation is inevitable \cite{chiani}. (3) With the adaptive
solution, the receiver filter is directed by the decisions made in
the previous time instance. Therefore, erroneous decisions lead to
unreliable filter weights.

To address { these problems}, an adaptive multiuser decision
feedback solution is proposed for time-varying multiple access MIMO
channels. The so-called adaptive decision feedback detection with
parallel interference cancellation and constellation constraints
(P-DFCC) algorithm proposes a constrained constellation map which
introduces a list which serves as the feedback candidates for P-DF
detection. By calculating the bit and symbol reliability, a higher
degree of freedom is introduced to refine the unreliable estimates
in the cancellation stage. The proposed algorithm is able to
significantly improve the performance for a traditional adaptive
S-DF or P-DF detector and close the gap from the MLD. Thanks to the
reliability calculation, the proposed algorithm obtains the
combination list at a small additional computational cost.

We also consider a spatially multiplexed multiuser iterative
detection and decoding (IDD) scheme incorporated with the proposed
structure. In this coded system, the soft-input soft-output (SISO)
detector is required to produce soft-decision values in terms of
log-likelihood ratio (LLR). The proposed SISO detector uses the
produced combination list to compute the likelihood of each
transmitted bit, the probability of the decision is conveyed. This
SISO detector is further concatenated with a SISO channel decoder to
form a turbo structure which allows a lower SNR requirement for the
adaptive MUD receiver. Computer simulations indicate that the {
proposed P-DFCC algorithm} significantly outperforms the
conventional S/P-DF schemes (i.e. \cite{choi} ) and approaches the
optimal performance with very low additional detection complexity.

The main contributions of this paper are:
{\begin{itemize}
\item An adaptive decision feedback based algorithm is developed for data detection in time-varying MIMO channels.
\item A P-DF receiver structure is investigated with the adaptive scheme, the constellation constraints (CC) is incorporated in the receiver to enhance the performance of interference cancellation.
\item The error performance and the detection complexity of the proposed algorithm are compared with several popular existing S/P-DF and optimal detection schemes.
\item A SISO detector is developed as a component of a multiuser IDD receiver structure.
\end{itemize}}

The organization of this paper is as follows. Section II gives the
multiuser spatial multiplexing MIMO system model as well as the
conventional S/P-DF { detector} and optimal detection criterion. The
proposed P-DFCC and its implementation are described in section III
and followed by a complexity comparison in section IV. The iterative
detection and decoding structure is introduced in Section V. The
simulation results are given in Section VI and Section VII concludes
the paper.

\section{System and Data model}

Let us consider a model of an uplink MU-MIMO system with $K$ UEs and
an access point (AP). Each UE is equipped with a single antenna. At
the receiver of the AP, $N_R$ receive antennas are available for
collecting and processing the signals. Throughout this paper, the
complex baseband notation is used while vectors and matrices are
written in lower-case and upper-case boldface, respectively. We
assume that the signals of the UEs are perfectly synchronized at the
AP, at each time instant $[i]$ $K$ users simultaneously transmit $K$
symbols which are organized into a vector ${\boldsymbol s}[i] =
\big[ s_1[i], ~s_2[i], ~ \ldots,~ s_{K}[i] \big]^T$, where
$(\cdot)^T$ denotes the transpose operation, and whose entries are
chosen from a complex $C$-ary constellation set $\mathcal{X} = \{
a_1,~a_2,~\ldots,~a_C \}$. The symbol vector ${\boldsymbol s}[i]$ is
transmitted over time-varying channels and the received signal is
processed by the receiver at the AP with $N_R$ spatially
uncorrelated antennas. The received signal is collected to form an
$N_R \times 1$ vector with sufficient statistics for detection
\begin{equation}\label{MIMO}
\begin{aligned}
{\boldsymbol r}[i] &= \sum_{k=1}^K{\boldsymbol h}_k[i]s_k[i] +
{\boldsymbol v}[i] \\
&= {\boldsymbol H}[i] {\boldsymbol s}[i] +
{\boldsymbol v}[i],
\end{aligned}
\end{equation}
where the $N_R \times 1$ vector ${\boldsymbol v}[i]$ represents a
zero mean complex circular symmetric Gaussian noise with covariance
matrix $E\big[ {\boldsymbol v}[i] {\boldsymbol v}^H[i] \big] =
\sigma_v^2 {\boldsymbol I}$, $\sigma_v^2$ is the noise variance and
${\boldsymbol I}$ is the identity matrix, $E[ \cdot]$ stands for the
expected value and $(\cdot)^H$ denotes the Hermitian operator. The
symbol vector ${\boldsymbol s}[i]$ has zero mean and a covariance
matrix $E\big[ {\boldsymbol s}[i] {\boldsymbol s}^H[i]\big] =
\sigma_s^2 {\boldsymbol I}$, where $\sigma_s^2$ is the signal power
for all transmitting UEs. Furthermore, the elements of
$\boldsymbol{H}[i]$ are the time-varying complex channel gains from
the $k$-th UE to the $n_R$-th receive antenna, which follow the
Jakes' model \cite{Jakes}. The $N_R \times 1$ vector $\boldsymbol
h_k[i]$ includes the channel coefficients of user $k$ such that
${\boldsymbol H}[i]$ is formed by the channel vectors of all users.
As the optimal SINR-based nulling and cancellation order (NCO)
\cite{choi} requires a high computational complexity, we determine
the NCO by computing the norms of the column vectors corresponding
to all users and we detect them in decreasing order of their norms.

\subsection{Optimal Detection}

\begin{figure}{
\centering \mbox{\epsfxsize=2.5in \epsffile{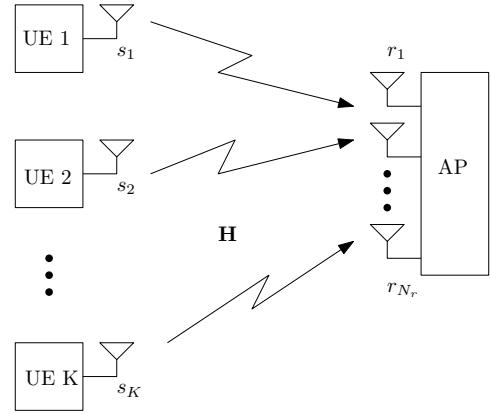}}
\caption{Spatially multiplexed multiple access system. We assume the
transmitted signal from $K$ UEs are spatially uncorrelated and $K
\leq N_R$. }\label{Fig:MAC}}
\end{figure}

The optimal ML detection algorithm tries all the possible transmitted signal vectors with the given channel $\boldsymbol{H}$, the detector computes the Euclidean distance by $\mathcal{J}(s)_{\scriptsize \mbox{Euclidean}} = \| \boldsymbol{r} - \boldsymbol{H}\hat{\boldsymbol{s}}  \| ^2$, the signal vector with the minimum Euclidean distance is determined as the estimate of the transmitted signal:
\begin{equation}\label{eq:ML}
\begin{aligned}
     \hat{\boldsymbol{s}}_{ML} & = \arg \max_{\boldsymbol{s} \in \mathcal{X}^{K}} P(\boldsymbol{r}|\boldsymbol{s}) \\
                  & = \arg \max_{\boldsymbol{s} \in \mathcal{X}^{K}} \frac{1}{({\pi\sigma_v^2})^{K}} \exp ( -\frac{\|\boldsymbol{r}-\boldsymbol{H}\boldsymbol{s}\|^2}{\sigma_v^2}   ) \\
                  & = \arg \min_{\boldsymbol{s} \in \mathcal{X}^{K}} \mathcal{J}(s)_{\scriptsize \mbox{Euclidean}},
\end{aligned}
\end{equation}
Similarly to MAP detection, the algorithm requires an exhaustive search of $|\mathcal{X}|^{K}$ equations in (\ref{eq:ML}) . The high complexity of the metric calculation prevents the actual application of these detectors in the real world, except for very small systems and constellations.

\subsection{Successive and Parallel DF Receivers}

\begin{figure*}{
\centering \mbox{\epsfxsize=5in \epsffile{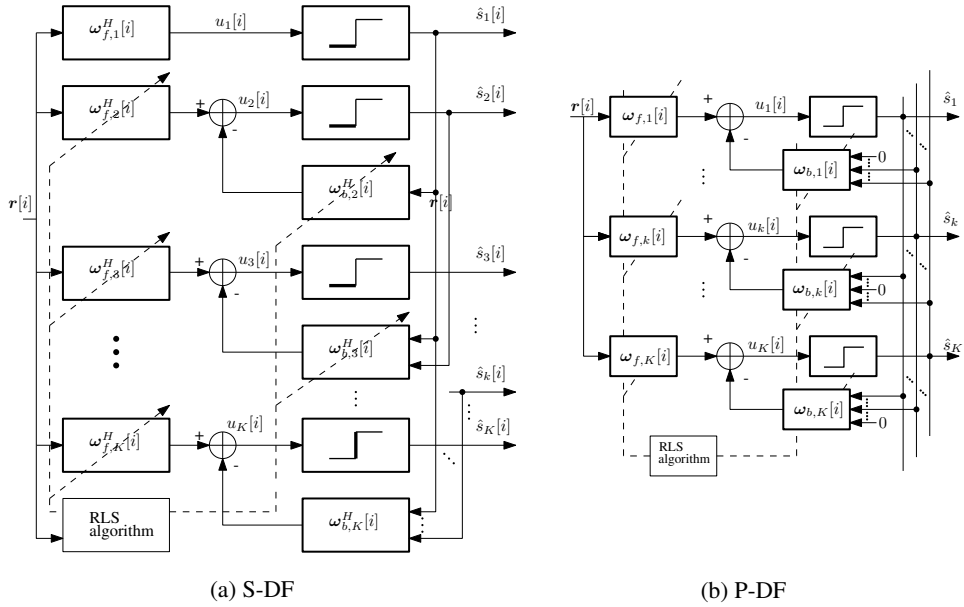}} \caption{{
Block diagram of the conventional (a) S-DF scheme and (b) P-DF
scheme. An RLS algorithm is employed to iteratively obtain the
filter weights.} }\label{Fig:S/P-DF}}
\end{figure*}

Let $\hat{\boldsymbol{s}}[i] =
\big{[}\hat{s}_1[i],\hat{s}_2[i],\ldots,\hat{s}_{K}[i]\big{]}^T$
represent the detected symbol {  vector. The soft symbol estimates
$u_k[i]$ are obtained by calculating the difference between the
output of the forward receive filter and the output of backward
receive filter as described in \cite{choi}} and given by
\begin{equation}\label{df}
u_k[i] = \boldsymbol{\omega}^H_{f,k}[i]\boldsymbol{r}[i] -
\boldsymbol{\omega}^H_{b,k}[i]{\boldsymbol{d}}_{k}[i],
\end{equation}
where {  the column vector $\boldsymbol{\omega}_{f,k}[i] \in
\mathbb{C}^{N_R \times 1}$} denotes the forward receive filter. The
{  column vector} $\boldsymbol{\omega}_{b,k}[i]$ indicates a {
backward} receive filter with the dimension $k-1$ {  for successive
decision feedback (S-DF) detection, or $K-1$ for parallel decision
feedback (P-DF) detection.} {

\subsubsection{S-DF}

S-DF detection is illustrated in Fig.\ref{Fig:S/P-DF}(a), where the
backward receive filter $\boldsymbol{\omega}_{b,k}[i] \in
\mathbb{C}^{k-1}$ has $k$ weight elements, and the size of
$\boldsymbol{\omega}_{b,k}[i]$ increases as $k$ raises. } The
forward filters $\boldsymbol{\omega}_{f,k}[i]$ act as the nulling
vectors of the V-BLAST algorithm. Then for each data stream $k = 1,
\ldots, K$, the decisions are accumulated and cancelled by the
$(k-1)$-dimensional filter $\boldsymbol{\omega}_{b,k}[i]$. {  The
backward receive filter is initialized by $\boldsymbol{\omega}_{b,1}
= \boldsymbol{0} $ for the first user. For the following users, the
$(k-1)$-dimensional detected symbol vector is obtained as}
\begin{equation}
{\boldsymbol{d}}_{k}[i]=\big{[}\hat{s}_1,\hat{s}_2,\ldots,\hat{s}_{k-1}\big{]}^T.
\end{equation}
{  The S-DF detection can provide a diversity order of $N_R - K + k$
for each user $k$ assuming that perfect interference cancellation is
performed by the receiver}.

{

\subsubsection{P-DF}

By assuming perfect interference cancellation, P-DF is able to
provide a higher diversity order compared to the S-DF based
detection algorithms. Similar to S-DF scheme, the P-DF first
processes the received signal $\boldsymbol{r}[i]$ by the forward
receive filter $\boldsymbol{\omega}_{f,k}[i] \in \mathbb{C}^{N_R
\times 1}$. However, as shown in Fig.\ref{Fig:S/P-DF}(b), the
backward receive filter is different from S-DF, which is given as
${\boldsymbol{\omega}}_{b,k}[i] \in \mathbb{C}^{(K-1) \times 1}$,
and the decision feedback symbol vector is defined as}
\begin{equation}
{\boldsymbol{d}}_{k}[i]=\big{[}\hat{s}_1,\ldots,\hat{s}_{k-1},0,\hat{s}_{k+1}\ldots,\hat{s}_{K}\big{]}^T.
\end{equation}
where the decisions for user $\hat{s}_k = \mbox{Q}\{ u_k[i] \}$ are
obtained by applying a slicer represented by $\mbox{Q}\{\cdot\}$.

For notational convenience, the forward and backward filters can be
concatenated together as \cite{choi}
\begin{equation}
\tilde{\boldsymbol{\omega}}_k[i] =
\begin{cases}
\boldsymbol{\omega}_{f,k}[i], &  k=1\\
\big{[}\boldsymbol{\omega}_{f,k}^T[i],\boldsymbol{\omega}_{b,k}^T[i]\big{]}^T,
&  k=2,\ldots, K.
\end{cases}
\end{equation}
The input can also be concatenated as
\begin{equation}\label{concat2}
\tilde{\boldsymbol{r}}_k[i] =
\begin{cases}
\boldsymbol{r}[i], &  k=1\\
\big{[}\boldsymbol{r}^T[i], {\boldsymbol{d}}_{k}^T[i]\big{]}^T,
&  k=2,\ldots, K.
\end{cases}
\end{equation}
Then, we can rewrite the soft estimates (\ref{df}) as
\begin{equation}\label{concatenated}
u_{k}[i] =
\tilde{\boldsymbol{\omega}}_k^H[i]\tilde{\boldsymbol{r}}_k[i].
\end{equation}
{  The forward and backward filters can be jointly optimized by
using an MMSE criterion or solving a lest squares problem. For the
sake of computational complexity, in the proposed structure the
recursive least squares (RLS) algorithm is adopted for the design of
the forward and backward filters. It should be noted that other
advanced parameter estimation algorithms such as reduced-rank
techniques \cite{mdfpic,jidf} can also be used.}

\section{Adaptive P-DF with Constellation Constraints}

\subsection{Computation of P-DF filters}

As a result, the structure and the signal processing model of the
proposed DF detector are depicted in Fig.{\ref{Fig:CC}}. We denote
the receive filter of each user as
$\tilde{\boldsymbol{\omega}}_k[i]$ ($k = 1, \ldots, K$), and the
value of each entry can be obtained by solving the standard least
squares (LS) problem. The LS cost function with an exponential
window is given by
\begin{equation}\label{lsp}
\mathcal{J}_k[i] = \sum_{\tau=1}^{i} \lambda^{i-\tau}
\Big{|}\hat{s}_k[\tau]-\tilde{\boldsymbol{\omega}}^H_k[i]\tilde{\boldsymbol{r}}_k[\tau]\Big{|}^2,
\end{equation}
where $0\ll\lambda<1$ is the forgetting factor, the scalar $\hat{s}_k[\tau]$ denotes the detected
signal in the time index $\tau$ or the known pilots where $\hat{s}_k[\tau] = s_k[\tau]$. The optimal tap
weight minimizing $\mathcal{J}_k[i]$ is given by
\begin{equation}\label{ls}
\tilde{\boldsymbol{\omega}}_k[i] =
\boldsymbol{\Phi}_k^{-1}[i]\boldsymbol{p}_k[i],
\end{equation}
where the time-averaged cross correlation matrix is obtained by
$\boldsymbol{\Phi}_k[i] = \sum_{\tau=1}^{i}
\lambda^{i-\tau}\tilde{\boldsymbol{r}}_k[\tau]\tilde{\boldsymbol{r}}_k^H[\tau]$
and $\boldsymbol{\Phi}_k[0] = \boldsymbol{0}$,
the time-averaged cross correlation vector is defined by
$\boldsymbol{p}_k[i] = \sum_{\tau=1}^{i}
\lambda^{i-\tau}\tilde{\boldsymbol{r}}_k[\tau]\hat{{s}}_k^*[\tau].$

Using the recursive least squares (RLS) algorithm \cite{book:H01}, the
optimal weights in (\ref{ls}) can be calculated recursively as
follows:
\begin{equation}\label{rls1}
\boldsymbol{q}_k[i]=\boldsymbol{\Phi}_k^{-1}[i-1]\boldsymbol{r}_{k}[i],
\end{equation}
\begin{equation}\label{rls2}
\boldsymbol{k}_k[i]=\frac{\lambda^{-1}\boldsymbol{q}_k[i]}{1+\lambda^{-1}\boldsymbol{r}_{k}^H[i]\boldsymbol{q}_{k}[i]},
\end{equation}
\begin{equation}\label{rls3}
\boldsymbol{\Phi}_k^{-1}[i]=\lambda^{-1}\boldsymbol{\Phi}_k^{-1}[i-1]-\lambda^{-1}\boldsymbol{k}_k[i]\boldsymbol{q}^H_k[i],
\end{equation}
\begin{equation}\label{rls4}
\tilde{\boldsymbol{\omega}}_{k}[i]=\tilde{\boldsymbol{\omega}}_{k}[i-1]+\boldsymbol{k}_k[i]\xi_k^*[i],\\
\end{equation}
where
\begin{equation}\label{rls5}
\xi_k[i]=
\begin{cases}
{s}_{k}[i] - \tilde{\boldsymbol{\omega}}_{k}^H[i-1]\tilde{\boldsymbol{r}}_{k}[i],   & \mbox{Training Mode,} \\
\hat {s}_{k}[i] - \tilde{\boldsymbol{\omega}}_{k}^H[i-1]\tilde{\boldsymbol{r}}_{k}[i],    & \mbox{Decision-directed Mode.}  \\
\end{cases}
\end{equation}
As indicated in (\ref{rls5}), this adaptive detection algorithm
works in two modes. The first one is employed with the training
sequence, while the second one is the decision-directed mode that is
switched on after the filter weights converge. In the
decision-directed mode, the mean square error (MSE) of the estimated symbols has a
major impact on the performance of adaptive DF algorithms. This is
because the detection error of the current user may propagate
throughout the detection of the following users. Moreover, in
time-varying channels a poor $\xi_k[i]$ can easily damage the
$\tilde{\boldsymbol{\omega}}_k[i]$ in equation ({\ref{rls4}})
resulting in burst errors.

\subsection{P-DF with Constellation Constraints}
In order to address this problem, the proposed P-DF with constellation constraints (P-DFCC) structure introduces a number of selected constellation points as the candidate decisions when the filter output $u_{k}[i]$ is determined unreliable. After the system is switched to the decision-directed mode, the concatenated filter output $u_k[i]$ is checked by the CC device which is illustrated in Fig. {\ref{Fig:CC}}. The CC structure is defined by the threshold distance $d_{\scriptsize \mbox{th}}$, {which can be a constant or determined in terms of SNR. The reliability of the estimated symbol is determined by the Euclidean distance between the symbol estimates and its nearest constellation points, which are given by
\begin{equation}\label{Euclidean}
d_{k} = \min_{a_c\in\mathcal{X}}\{|u_{k}[i]-a_c|\},
\end{equation}
where $a_c$ denotes the constellation point which is the nearest to the soft estimation $u_{k}[i]$ of the $k$-th symbol. The CC device distinguishes the reliable estimation from the unreliable ones, which allows the P-DFCC to avoid redundant processing with reliable feedbacks and maintain the complexity at the same level of the conventional P-DF structure. The following is devoted to describe the detection of $\hat{s}_{k}[i]$ for the $k$-th user.

\begin{figure}{
\centering \mbox{\epsfxsize=2.5in \epsffile{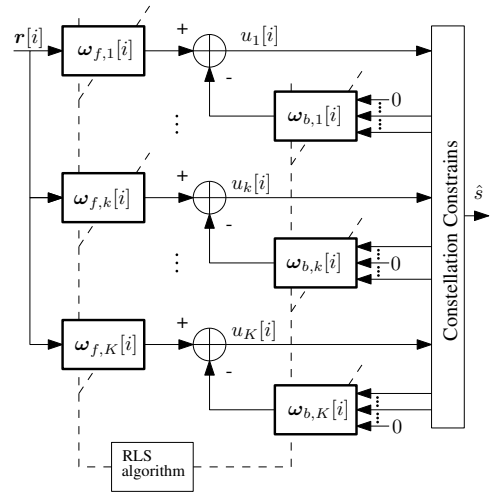}} \caption{{
Block diagram of the proposed P-DFCC multi-user detector. There are
$K-1$ interference symbols for each user's backward filter. }
}\label{Fig:CC}}
\end{figure}

 \begin{figure}{
 \centering \mbox{\epsfxsize=2.5in
 \epsffile{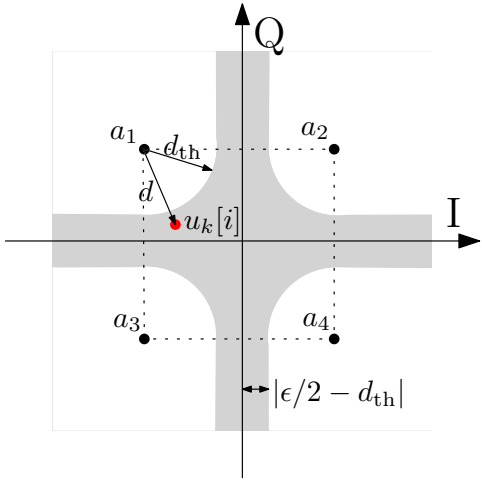}}
 \caption{The constellation constraints (CC) device. The CC procedure is invoked as the soft estimates $u_k[i]$ dropped into the shaded area. Parameter $\epsilon$ denotes the distance between 2 nearest constellation symbols.}\label{cc}}
 \end{figure}
%

Let us define two regions for the QPSK constellation map: (1) The region inside the square obtained by connecting four $a_c$, the $a_c$ are assumed to have the form, $a_c = \Big{(}\pm{{\epsilon}/{2}}, \pm ({\epsilon}/2)j\Big{)}$, where $\epsilon$ is the distance between two nearest constellation symbols. The estimate $u_k[i]$ is considered inside the square if the following equations hold
\begin{equation}\label{sa1}
\begin{cases}
{\big{|}\Re\{u_k[i]\}\big{|}} \leq {\epsilon}/{{2}} \\
{\big{|}\Im\{u_k[i]\}\big{|}} \leq {\epsilon}/{{2}}. \\
\end{cases}
\end{equation}
where $\Re\{ \cdot \}$ and $\Im\{ \cdot \}$ denote the real part and the imaginary part of a complex-valued quantity, respectively.

(2) Otherwise, the estimate is in the region outside the square obtained.

\subsubsection{CASE 1 inside the square}
In the first case, the estimate $u_k[i]$ is considered as unreliable if the following equation holds
\begin{equation}
d_k > d_{\scriptsize \mbox{th}}. \\
\end{equation}
where $d_k$ denotes the distance between the estimated symbol $u_k[i]$ and its nearest constellation point and $a_c$ is each element of the constellation points. Otherwise, the estimated symbol is closer to the constellations and the decision is considered as reliable.

\subsubsection{CASE 2 outside the square}
In this case, the equations of (\ref{sa1}) do not hold, where the estimated symbol $u_k[i]$ is outside the square. In this case, the decision is determined unreliable if the distance from $u_k[i]$ to I(In-phase)-axis and Q(quadrature)-axis is small. Therefore, the estimate is unreliable if any of the following equations holds

\begin{equation}\label{case2re}
 {\big{|}{\Re}\{u_k[i]\}\big{|}} < {\epsilon}/{2} - d_{\scriptsize \mbox{th}},
\end{equation}

\begin{equation}\label{case2im}
 {\big{|}{\Im}\{u_k[i]\}\big{|}} < {\epsilon}/{2} - d_{\scriptsize \mbox{th}}.
\end{equation}
Otherwise, the estimated symbol is far away from the axis borders and the estimate is considered as reliable.

This can be further extended to multi-tier constellations (eg.16-QAM) where the outer-tier would be similar to CASE 2, but we should also include two additional equations in addition to (\ref{case2re}) and (\ref{case2im}) which are given as
\begin{equation}\label{case2re_16qam}
 \min {\big{|}{\Re}\{u_k[i]\} \pm \epsilon \big{|}} < {\epsilon}/{2} - d_{\scriptsize \mbox{th}},
\end{equation}
\begin{equation}\label{case2im_16qam}
 \min {\big{|}{\Im}\{u_k[i]\} \pm \epsilon \big{|}} < {\epsilon}/{2} - d_{\scriptsize \mbox{th}}.
\end{equation}
where $\big{|}{\Re}\{u_k[i]\} \pm \epsilon \big{|}$ are the distances between $u_k[i]$ and two vertical lines across the points $(0,\pm \epsilon)$, respectively. The matrices $\big{|}{\Im}\{u_k[i]\} \pm \epsilon \big{|}$ are similarly defined as the distances between $u_k[i]$ and two horizontal lines across points $(\pm \epsilon ,0)$, respectively. Therefore, for 16-QAM constellations, the estimate is considered as unreliable if any one of the four equations above (\ref{case2re}-\ref{case2im_16qam}) holds. On the other hand, for the inner-tier constellations, if
\begin{equation}
 \min{\big|}a_k[i] - u_k[i]{\big|} \geq d_{\scriptsize \mbox{th}} \qquad \forall c,
\end{equation}
was true, the estimate is considered as unreliable. The CC device distinguishes the reliable feedback signals from the unreliable ones, which allows the P-DFCC to maintain the complexity at the same level of the conventional DF structure.}

\subsubsection*{Reliable}
If the filter output $u_k[i]$ is dropped into the lighted area of the constellation map, the decision is considered reliable. The tentative decision of $s_k[i]$ is obtained by
\begin{equation}
 \mathcal{L}_k[i] =  \arg_{a_c} \min{\big|}a_c - u_k[i]{\big|}
\end{equation}

\subsubsection*{Unreliable}
If it is the case that $u_k[i]$ is determined unreliable, we proceed by organizing the Euclidean distance obtained by (\ref{Euclidean}) in decreasing order, a list of tentative decisions of $s_k[i]$ is obtained as given by
\begin{equation}\label{list_un}
 \mathcal{L}_k[i] \triangleq \{c_1, c_2, \ldots, c_\tau\}_k,
\end{equation}
where $1\leq \tau \leq |\mathcal{X}|$, and
\begin{equation}
 \mbox{de}[c_1] \leq \mbox{de}[c_2] \leq \ldots, \mbox{de}[c_\tau],
\end{equation}
where $\mbox{de}[\cdot]$ denotes the Euclidean distances between $u_k$ and $c_\tau$.

Therefore, for each user we obtain a tentative decision list $\mathcal{L}_k$. By listing all the combinations of the elements across $K$ users, a length $\varGamma$ tentative decision list is formed. {Each column vector on the list denotes a possible transmission symbol vector $\boldsymbol{s}'_l$ where $l = 1, \ldots, \varGamma$.} The size of the list is obtained by
\begin{equation}
 \varGamma = \prod_{k=1}^{K}|\mathcal{L}_k|,  \qquad 1\leq\varGamma\ll |\mathcal{X}|^K,
\end{equation}
where $|\cdot|$ denotes cardinality. In order to obtain an improved performance, the maximum likelihood (ML) rule can be used to select the best among the $\varGamma$ candidate symbol vectors. The cost function for the ML selection criterion, which is equivalent to the minimum Euclidean distance criterion and the selected vector is given by
\begin{equation}\label{Equ:ML}
 \boldsymbol{s}'_{\scriptsize \mbox{ML}} = \arg\min_{l = 1,\ldots,\varGamma} \big{\|}\boldsymbol{r}[i] - \boldsymbol{H}\boldsymbol{s}_l'[i]\big{\|}^2,
\end{equation}
where $\boldsymbol{s}'_{\scriptsize \mbox{ML}}$ is the ML selected vector, which can be used as the feedback symbols as well as the decision vector.

The number of $\varGamma$ could be considered as a reflection of the
trade-off between complexity and performance. By assuming a large
threshold $d_{\scriptsize \mbox{th}}$, the proposed scheme is able
to tolerate a higher error energy and results in a smaller
$\varGamma$ but suffers from a performance loss. For an extreme case
where $d_{\scriptsize \mbox{th}} = \inf$, the proposed detector is
equivalent to a conventional D-DF detector. {  In contrast, if
$d_{\scriptsize \mbox{th}} = 0$, we have $\varGamma =
|\mathcal{X}|^K$ which means the proposed receiver performs DF
detection with an ML rule that allows the search for an ML solution
for each user.  It is also worth to mention that a maximum
${\tau_{\scriptsize \mbox{max}}}$ can be set to guarantee
$1\leq\varGamma\ll |\mathcal{X}|^K$, which prevent high complexity
in very low SNR range.

By introducing a constellation constraint, (a) the detection
diversity is directly related to the threshold $d_{\scriptsize
\mbox{th}}$: a decrease in the value of $d_{\scriptsize \mbox{th}}$
could result in a longer list which may increase the diversity
order. (b) In the low SNR region, it is also likely to obtain a
longer list than that in a high SNR region, hence the diversity
order tends to be higher. On the other hand, for the high SNR
region, all the symbol estimates are considered reliable and the
diversity order tends to be the same of a conventional P-DF (this is
similar to increasing the threshold $d_{\scriptsize \mbox{th}}$).
This implies that the gain provided by P-DFCC is higher for a small
to medium region of SNR.}

\subsection{Channel Estimation}

As we discussed in the previous sections, the MIMO channel state information is required for the ML rule (\ref{Equ:ML}) and for generating the cancellation ordering codebook for ordered processing \cite{FL11}. The LS channel estimation algorithm has been investigated in \cite{WLM11}. Based on a weighted average of error squares, the estimated channel minimizes the cost function whose expression at time instant $i$ is defined as
\begin{equation}
\mathcal{J}_{\hat{\boldsymbol{H}}}[i] = \sum_{\tau=1}^{i} \lambda^{i-\tau} \Big{|} \boldsymbol{r}[\tau]-\hat{\boldsymbol{H}}[i]\boldsymbol{s}[\tau] \Big{|}^2,
\end{equation}
where $\hat{\boldsymbol{H}}[i]$ is the channel matrix estimate at time instant $i$. The quantities $\boldsymbol{r}[\tau]$ and $\boldsymbol{s}[\tau]$ are the received signal and the pilot symbol vectors at the time instant $\tau$, respectively.

To minimise the cost function, the gradient of the cost function with regard to the estimated channel matrix should be equated to a zero matrix as
\begin{equation}
 \nabla _{\hat{\boldsymbol{H}}[i]}\mathcal{J}_{\hat{\boldsymbol{H}}}[i] = \boldsymbol{0}_{N_R,K}.
\end{equation}
By solving the above equation, the LS estimate of the channel matrix is obtained as
\begin{equation}
\begin{aligned}
{\hat{\boldsymbol{H}}}[i]  & = \Big{(}\sum_{\tau=1}^{i} \lambda^{i-\tau}\boldsymbol{r}[\tau]\boldsymbol{s}^H[\tau] \Big{)} \Big{(}\sum_{\tau=1}^{i} \lambda^{i-\tau}\boldsymbol{s}[\tau]\boldsymbol{s}^H[\tau] \Big{)}^{-1} \\
& = \boldsymbol{D}[i] \boldsymbol{\Phi}^{-1}[i].
\end{aligned}
\end{equation}
In order to avoid the matrix inversion operation $(\cdot)^{-1}$, a recursive algorithm is developed. Let us define
\begin{equation}
{\boldsymbol{\Phi}}^{-1}[i] = \boldsymbol{P}[i],
\end{equation}
where $\boldsymbol{D}[i]$ can be obtained iteratively by
\begin{equation}
\boldsymbol{D}[i] = \lambda \boldsymbol{D}[i-1] + \boldsymbol{r}[i]\boldsymbol{s}[i]^H
\end{equation}
and $\boldsymbol{P}[i]$ is calculated iteratively by using the matrix inversion lemma,
\begin{equation}
\boldsymbol{P}[i] = \lambda^{-1} \boldsymbol{P}[i-1] - \frac{\lambda^{-2}\boldsymbol{P}[i-1]\boldsymbol{s}[i]\boldsymbol{s}[i]^H\boldsymbol{P}[i-1]}{1+\lambda^{-1}\boldsymbol{s}[i]^H\boldsymbol{P}[i-1]\boldsymbol{s}[i]}.
\end{equation}
The initial state of the parameters are set as $\boldsymbol{D}[0]=\boldsymbol{0}_{N_R,K}$ and $\boldsymbol{P}[0]=\delta_c^{-1}\boldsymbol{I}$, where $\delta_c$ is a small constant.

\section{Iterative Detection and Decoding}

In the previous section, we have introduced the concept of constellation constraints and its implication for an uncoded multi-user detection algorithm. In order to reduce the SNR requirement for a MIMO receiver, error-control coding is essential for the system. Iterative detection and decoding (IDD) has been recognized as central technique for solving a large number of decoding and detection problems in wireless communications. In this section, the we are interested in developing IDD algorithms for spatially multiplexed multi-user data streams.

For a multi-user MIMO IDD transmission system, the message is first encoded by an encoder, the coded bits are then interleaved and the coded bits are mapped to symbols before radiating from a transmitting antenna. At the receiver side, the P-DFCC detector is applied to detect the transmitted symbols and convert the symbol probability to bit probability in the form of LLRs. The \textit{extrinsic information} $L_e(\cdot)$ is then exchanged between the detector and the decoder with several iterations. The \textit{a posteriori} probability of the transmitted bits are then finally obtained at the output of the decoder.

On one hand, the encoder and decoder blocks are considered as the
outer code of a serially concatenated structure, when a
non-systematic convolutional coded (NSC) is applied, the BCJR
\cite{BCJR74} based MAP or log-MAP decoding algorithm can be applied
as well as the lower complexity alternative named soft-output
Viterbi algorithms (SOVA) \cite{SOVA}. Instead of using a
convolutional code as the channel code, turbo codes and LDPC
\cite{UHL11} codes along with advanced decoding algorithms
\cite{vfap} can also be used in this structure to obtain a
near-capacity performance \cite{HB03} \cite{VHK04}. On the other
hand, the mapping and MIMO detection blocks are considered as the
inner component of the serially concatenated structure. In general,
MAP is the optimal algorithm used as the SISO detection component in
the IDD receiver. The MAP detector provide the optimal BER
performance, however, the complexity is extreme. In order to solve
this problem, a ``list" version of SD was developed by Hochwald and
ten Brink without significant loss of performance \cite{HB03}. The
complexity of the MIMO detection is further brought down by
introducing soft parallel interference cancellation (SPIC) in
\cite{WP99}, \cite{WP99} at the cost of a performance loss. In this
section, we adapt the proposed P-DFCC detection algorithm into the
IDD structure.

In the coded systems, the model in (\ref{MIMO}) is used repeatedly to describe transmit streams of data bits which are separated into blocks. For a given block, the symbol vector $\boldsymbol{s}$ is obtained by mapping $\boldsymbol{b} = [b_1,..., b_j,..., b_{K{\cdot}{J}}]$ coded bits. The quantity $J$ is the number of bits per constellation symbol. For coded transmissions, the vector $\boldsymbol{b}$ is designated as the output of a forward error-correction code of rate $R < 1$ that introduces redundancy. The transmission rate is then $RKJ$ bits perreceived vector. In the IDD processing, the detector makes decisions by using the knowledge of correlations across time instants $[i],i = 0, 1, \ldots, I$ provided by the channel decoder, and the channel decoder needs to decode the bit information by using the likelihood information on all blocks obtained from the soft output detector.

For each user, a block of received signals $\boldsymbol{r}[i]$ is used to compute the \textit{a posteriori} probability in the form of log-likelihood-ratios (LLRs), with P-DF, the MIMO input-output relation (\ref{MIMO}) has been transformed in to $K$ parallel data streams. By assuming these $K$ streams are statistically independent, we may approximate the \textit{intrinsic a posteriori} LLRs as \cite{thesis_studer}
\begin{equation}
 \Lambda_1^p[b_{j,k}[i]] \approx \log {\frac{P[b_{j,k}[i] = +1 | u_{k}[i] ]}{P[b_{j,k}[i] = -1 | u_{k}[i]]}} \qquad \forall j,k,
\end{equation}
where the equation can be solved by using Bayes' theorem and we leave the details to the references \cite{WP99,HB03}. We denote the \textit{intrinsic} information provided by the decoder as $\Lambda_2^p[b_{j,k}[i]]$ and the bit probability is obtained as
\begin{equation}
 P[b_{j,k}[i]] = \log\frac{P[b_{j,k}[i] = +1]}{P[b_{j,k}[i] = -1]} \qquad \forall j,k.
\end{equation}
From \cite{WP99}, the bit-wise probability is obtained by
\begin{equation}\label{bitp}
\begin{aligned}
 P[b_{j,k}[i] = \bar{b}_j] & = \frac{\exp\Big{(}{\bar{b}_j}\Lambda_2^p[b_{j,k}[i]]\Big{)}}{1+\exp\Big{(}{\bar{b}_j}\Lambda_2^p[b_{j,k}[i]]\Big{)}},\\
              & = \frac{1}{2}\left[1+\bar{b}_j\tanh\Big{(}\frac{1}{2}\Lambda_2^p[b_{j,k}[i]\Big{)}\right].
\end{aligned}
\end{equation}
where $\bar{b}_j = \{ +1, -1\}$.
Let us simplify the notation $P\big[s_k[i]\big] := P\big[s_k[i] = {c}_q\big]$ where $c_q$ is an element chosen from the constellation $\mathcal{X} = \{c_1,\ldots,c_q,\ldots,c_A\}$. The symbol probability $P[s_k[i]]$ is obtained from the corresponding bit-wise probability, and assuming the bits are statistically independent, we have
\begin{equation}\label{sybp}
\begin{aligned}
 P\big[s_k[i]\big] & = \prod_{j = 1}^{J} P\big[b_{j,k}[i] = \bar{b}_j\big],\\
          & = \frac{1}{2^J} \prod_{j = 1}^{J} \left[1+\bar{b}_j\tanh\Big{(}\frac{1}{2}\Lambda_2^p[b_{j,k}[i]\big{]}\Big{)}\right].
\end{aligned}
\end{equation}
From (\ref{bitp}) and (\ref{sybp}) we can conclude that $\sum_{|\mathcal{X}|} P\big[s_k[i]\big] = 1$. By organizing the probabilities obtained by (\ref{sybp}) in decreasing order of values, a list of tentative decisions of $s_k[i]$ is obtained in each stream as given by
\begin{equation}\label{list_co}
 \mathcal{L}_k^{\scriptsize \mbox{IDD}}[i] \triangleq \{c_1, c_2, \ldots, c_\tau\}_k,
\end{equation}
where $1\leq \tau \leq |\mathcal{X}|$ and
\begin{equation}
 Pr[c_1] \geq Pr[c_2] \geq \ldots, Pr[c_\tau],
\end{equation}
and
\begin{equation}
 Pr{[} c_q] \triangleq P\big{[}s_k[i] = c_q\big{|}{u_k}\big{]}.
\end{equation}
For the IDD coded structure we replace (\ref{list_un}) with
(\ref{list_co}), thanks to the error correction, for a moderate SNR,
the size of $\mathcal{L}_k^{\scriptsize \mbox{IDD}}$ is
significantly smaller than that value in (\ref{list_co}). The
pseudo-code for implementing the proposed P-DFCC with IDD structure
is detailed in Algorithm. \ref{alg:log-max-map-dfcc}.

\begin{algorithm}
\caption{Algorithm soft-output log-Max-DFCC Detection}
\label{alg:log-max-map-dfcc}
\algsetup{
linenosize=\small,
linenodelimiter=.
}
\begin{algorithmic}[1]
\REQUIRE{$\boldsymbol{r} \in \mathcal{C}^{N_R \times 1}$, $\boldsymbol{H} \in \mathcal{C}^{N_R \times K}$, constellation set $\mathcal{A}$, $\sigma_v^2$, $n \leftarrow 0$, $\boldsymbol{L}(b_{k,j}^{p1})$, $TI$. }

\STATE {Find the set of symbol vectors $\mathcal{X}_{k,j}^1\cap \mathcal{L}_k^{\scriptsize \mbox{IDD}}  $ and $\mathcal{X}_{k,j}^0\cap \mathcal{L}_k^{\scriptsize \mbox{IDD}}$}
\FOR{$lo \leftarrow TI$ \COMMENT{Turbo Iteration}}
   \FOR{$k \leftarrow 1,\ldots,K$}
     \FOR{$j \leftarrow 1,\ldots,J$}
    \FOR{${s} \in \mathcal{X}_{k,j}^1\cap \mathcal{L}_k^{\scriptsize \mbox{IDD}}$}
        \STATE{$\boldsymbol{b} \leftarrow \mbox{demap}({s})$, $\boldsymbol{b}_{k,j}\leftarrow 0$}
        \STATE{$\mbox{P}(\boldsymbol{x}) \leftarrow \frac{1}{2}(2\boldsymbol{b}_{[k,j]}-1)\boldsymbol{L}(b_{k,j}^{(p1)})$} \qquad \COMMENT{\textit{Symbol probability}}
        \STATE{$\lambda_n^1 \leftarrow \ln \mbox{P}(\boldsymbol{x}) - \frac{\|\boldsymbol{r}-\boldsymbol{Hs}\|^2}{\sigma_v^2}$}
    \ENDFOR

    \FOR{${s} \in \mathcal{X}_{k,j}^0\cap \mathcal{L}_k^{\scriptsize \mbox{IDD}}$}
        \STATE{$\boldsymbol{b} \leftarrow \mbox{demap}({s})$, $\boldsymbol{b}_{k,j}\leftarrow 0$}
        \STATE{$\mbox{P}(\boldsymbol{x}) \leftarrow \frac{1}{2}(2\boldsymbol{b}_{[k,j]}-1)\boldsymbol{L}(b_{k,j}^{(p1)})$} \qquad \COMMENT{\textit{Symbol probability}}
        \STATE{$\lambda_n^0 \leftarrow \ln \mbox{P}(\boldsymbol{x}) -  \frac{\|\boldsymbol{r}-\boldsymbol{Hs}\|^2}{\sigma_v^2}$}
    \ENDFOR
    \STATE{$L(b_{k,j}^{(e1)}) \leftarrow \max\{\lambda_n^1, n = 1,\ldots,|\mathcal{X}_{k,j}^1|\} - \max\{\lambda_n^0, n = 1,\ldots,|\mathcal{X}_{k,j}^0|\}$}
     \ENDFOR \qquad \qquad  \COMMENT{Antenna stream}
   \ENDFOR \qquad \qquad \COMMENT {\textit{Bit Label}}
   \STATE Deinterleave \textit{extrinsic} $\boldsymbol{L}(b_{k,j}^{(e1)})$
   \STATE Perform BCJR decoding and compute $\boldsymbol{L}(b_{k,j}^{(e2)})$
   \STATE Interleaving \textit{extrinsic} $\boldsymbol{L}(b_{k,j}^{(e2)})$ and feedback to detector.
\ENDFOR \qquad \qquad \qquad \qquad \qquad \COMMENT{\textit{Turbo Iteration}}
   \STATE Decision of systematic bit is obtained via $\mbox{sign} \{\boldsymbol{L}(m_k)\}$
\end{algorithmic}
\end{algorithm}

\section{Simulations}

In this section, several numerical examples are given to demonstrate
the overall system performance of using our algorithms. In the
following simulations, unless otherwise stated, we consider that the
proposed algorithms and all their counterparts operate with a
channel with independent and identically-distributed (i.i.d) block
fading model. The channel model is of Rayleigh random fading and the
coefficients are taken from complex Gaussian random variables with
zero mean and unit variance. Other parameters are also assumed: QPSK
is used; The transmitted vectors $\boldsymbol{s}[i]$ are grouped
into frames consisting of 500 vectors where the first
$\boldsymbol{s}[1],\ldots,\boldsymbol{s}[10]$ vectors are training
vectors. In each frame, the channel between a transmit and receive
antenna pair is fixed and a single path is assumed.

\begin{figure}[!htb]{
\centering \mbox{\epsfxsize=3.5in \epsffile{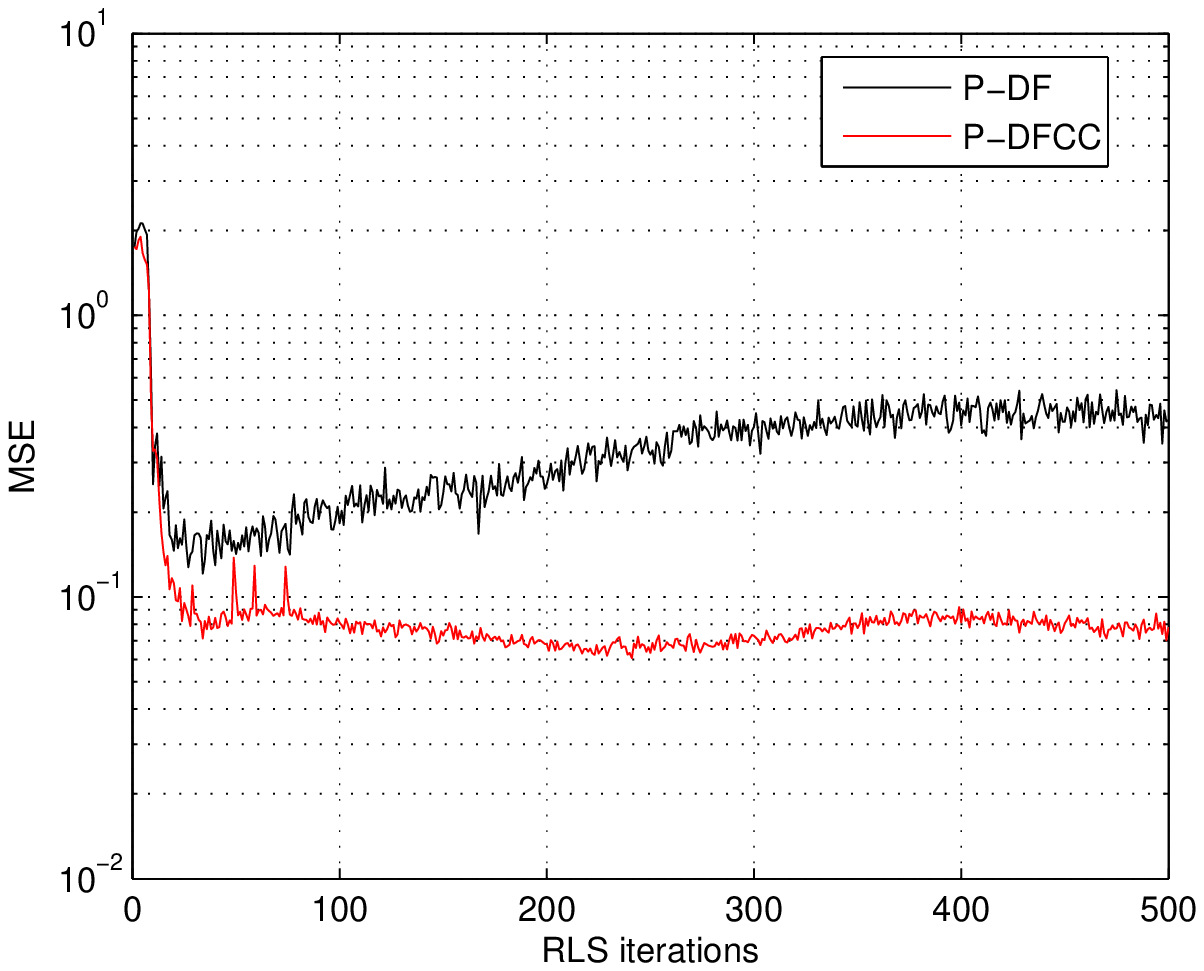}} \caption{MSE
of the estimated symbols in terms of RLS iterations, with $8$ users.
After 10 training vectors transmitted, the decision-directed mode is
switched on. The MSE is significantly reduced. }\label{Fig:MSEfdT}}
\end{figure}

Fig. {\ref{Fig:MSEfdT}} demonstrates the MSE for the symbol
estimation across all 8 user streams in terms of RLS iterations with
8 receiver  antennas configuration. $E_b /N_0 = 20 dB$, and the
normalized Doppler frequency $f_dT$ equals to $10^{-3}$. The
proposed P-DFCC scheme shows the improvement in terms of MSE. From
the figure we can see that the P-DFCC has the ability to track the
fading channel with  $f_dT = 10^{-3}$.

The performance is also measured in terms of bit error rate (BER),
obtained by $10^{4}$ Monte Carlo runs. In our simulations, the SNR
per transmitted information bit is defined as
\begin{equation}
\frac{E_b}{N_0}\Big{|}_{\mbox{\scriptsize dB}} = 10\log_{10}\Big{(}\frac{N_R}{R\log_2C}\cdot \frac{\sigma_s^2}{\sigma_v^2}\Big{)}.
\end{equation}

The total transmitted power $E_s = K \cdot \sigma_s^2$ which is
evenly distributed across $K$ active users. The $N_R$ receive
antennas collect a total power of $N_RE_s$ which carries
$K\log_{2}C$ coded bits or $RK\log_{2}C$ information bits. $R\leq1$
is the channel coding rate which introduces information redundancy.
The coding rate $R = 1$ is assumed for the simulations without
channel coding.

\begin{figure}[!htb]{
\centering \mbox{\epsfxsize=3.5in \epsffile{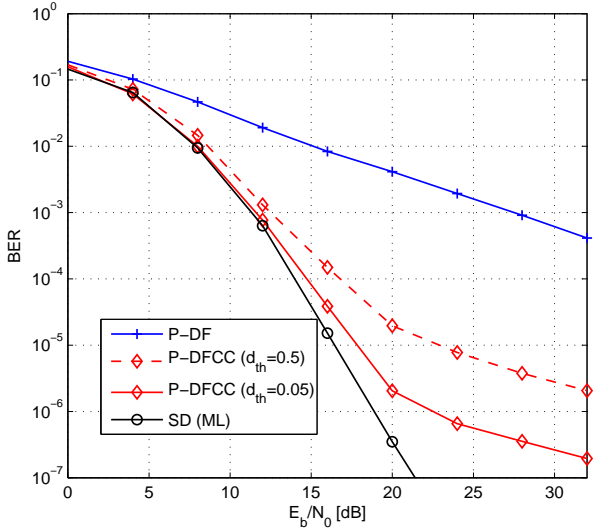}} \caption{BER
vs. $E_b/N_0$, the proposed P-DFCC detection achieves a near optimal
performance in a 4  user system configuration. The constellation
threshold $d_{\scriptsize \mbox{th}}$ introduces a trade off between
the performance and the complexity.}\label{Fig:BER}}
\end{figure}

Fig.\ref{Fig:BER} shows the BER against $E_b/N_0$. The channel  is
estimated by LS algorithms, the P-DF-RLS detector ($\lambda =
0.998$) proposed in \cite{choi} exhibits  about $7$dB performance
loss when the target BER equals $10^{-3}$ compared with the
performance of SD. As for the SD, with a sufficiently large sphere
radius selected, the SD can always produce an ML solution. With the
constellation constraint threshold $d_{\scriptsize \mbox{th}}=0.05$,
the proposed P-DFCC-RLS ($\lambda = 0.998$) algorithm shows a
near-optimal BER performance at the target BER equal to $10^{-3}$. {
From Fig.\ref{Fig:BER}, we can verify that the optimal ML detector
(or sphere decoder) is able to attain full diversity. On the other
hand, similar to an MMSE-based successive interference cancellation
(SIC) detection, the S-DF detector is able to obtain a diversity
order of $N_R-K + k$ and the BER performance is bounded by the user
with the worst performance.

It is also worth to mention that the diversity order of the
traditional P-DF algorithm is usually lower than the channel power
sorted S-DF \cite{FL11}, this is due to the problem of error
propagation. In P-DF, an erroneous symbol would propagate through
all other user's data stream. However, if all the detected symbols
are highly reliable, P-DF may provide a higher diversity order than
S-DF, this can be verified by assuming a perfect cancellation
scenario, where P-DF achieves full receive diversity order while
S-DF has only $N_R-K + k$.

By introducing a reliability checking procedure, the diversity order
of the proposed P-DFCC can be adjusted. The control of the diversity
order is twofold: (1) the selection of $d_{\scriptsize \mbox{th}}$.
From Fig.\ref{Fig:BER} we can see that the diversity order is
directly related to the threshold $d_{\scriptsize \mbox{th}}$.
Namely, decrease the value of $d_{\scriptsize \mbox{th}}$ could
change the shape of the constellation constraint and increase the
diversity order. (2) The received SNR region. In the low SNR region,
the scheme is likely to list a higher number of candidates than
those generated in a high SNR region and the performance approaches
the ML detector. On the other hand, for the high SNR region, all the
symbol estimates are considered reliable and the diversity order
tends to be the same of a conventional P-DF. Therefore, for the
proposed P-DFCC scheme the gain is higher for a small to medium
region of SNR.}

\begin{figure}[!htb]{
\centering \mbox{\epsfxsize=3.5in \epsffile{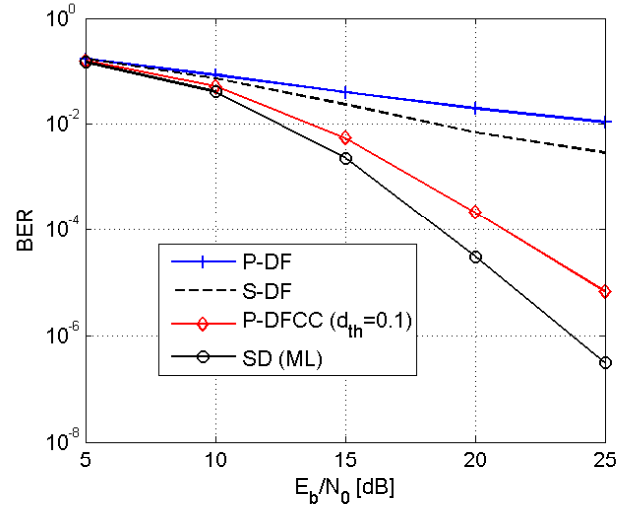}} \caption{BER
vs. $E_b/N_0$, the proposed P-DFCC detection achieves a near optimal
performance in a 4-user system configuration with 16-QAM symbols
}\label{Fig:BER_QAM}}
\end{figure}

Another simulation is carried out with 16-QAM symbols. The SNR against BER curves are plotted in Fig.\ref{Fig:BER_QAM}. The threshold is set to $d_{\scriptsize \mbox{th}} = 0.1$. With QPSK modulation the proposed P-DFCC detection algorithm is able to achieve a better performance compared with traditional P-DF as well as S-DF algorithms.

\begin{figure}[!htb]{
\centering \mbox{\epsfxsize=3.5in \epsffile{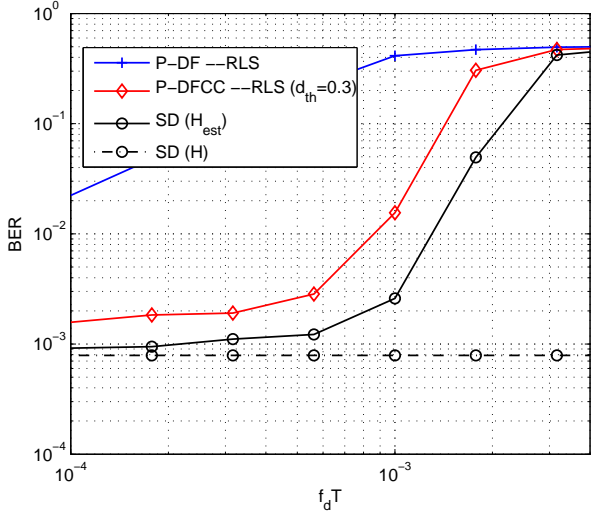}}
\caption{Comparison of BER performance for various normalized
Doppler frequency $f_dT$ when $K=4$ and $E_b/N_0$ =
14dB.}\label{Fig:fdT}}
\end{figure}

 Fig.{\ref{Fig:fdT}} presents the comparison of BER performance for various normalized Doppler frequency $f_dT$ (in the time-varying channels) when $E_b/N_0 = 14$ dB. In this simulation, each channel between a transmit and receive antenna pair varies accodrding to the Jakes' model \cite{Jakes}. LS channel estimation is applied to the unknown channel. The length of the training sequence is $I = 20$. The simulation results show that the proposed P-DFCC significantly improves the traditional P-DF detector and approaches the SD performance in time-varying channels.

\begin{figure}[!htb]{
\centering \mbox{\epsfxsize=3.5in \epsffile{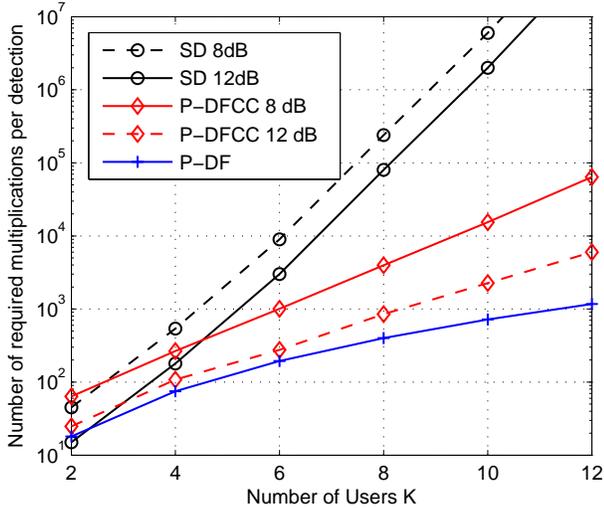}}
\caption{Complexity in terms of arithmetic operations against the
transmit antennas, the P-DFCC has a comparable complexity with P-DF
algorithm. $d_{\scriptsize \mbox{th}} =
0.3$.}\label{Fig:complexity}}
\end{figure}

In Fig.\ref{Fig:complexity}, the complexity is given by counting the required complex multiplications as the number of users increases.  P-DFCC has a complexity slightly above the P-DF while it achieves a significant performance improvement. The threshold $d_{\scriptsize \mbox{th}}$ is introduced to reduce the complexity and improve the performance. We use fixed complexity sphere decoders (FSD) \cite{BT08} to compare the complexity. It should be noted that FSD is one of the lowest complexity SD algorithms that are known.

\begin{figure}[!htb]{
\centering \mbox{\epsfxsize=3.5in \epsffile{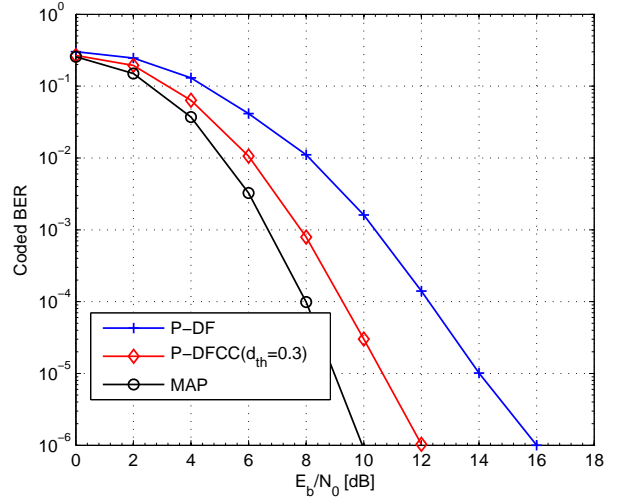}}
\caption{Coded BER curves of QPSK over $4 \times 4$ MIMO channels;
block size 1000 message bits, code rate $R = 1/2$, memory 2
convolutional code. }\label{Fig:CODED}}
\end{figure}

The curves in Fig.{\ref{Fig:CODED}} are given for convolutionally
coded BER performance on a Rayleigh block fading channel. The
proposed P-DFCC with $d_{th}=0.3$  improves the conventional P-DF
detection performance about 3 dB at the target coded BER equals to
$10^{-4}$. The P-DFCC detector approaches the optimal MAP detection
performance with only 1.5 dB performance loss when coded BER =
$10^{-4}$.

\section{Conclusion}

In this  paper, we have derived an adaptive decision feedback based
detector for MIMO transmission systems with varying channels. In
this context, we have presented a novel way to improve the BER
performance by using the parallel decision feedback with
constellation constraints approach, a threshold is introduced to
reduce the complexity and improve the performance. This approach has
the ability to reduce the MSE of traditional parallel decision
feedback detection, effectively improve the BER performance of
parallel interference cancellation schemes and obtain a close to
optimal performance with a low additional detection complexity.


%




\ifCLASSOPTIONcaptionsoff
  \newpage
\fi

\end{document}